\documentclass{article}

\usepackage{spconf,amsmath,graphicx}
\usepackage{graphicx}
\usepackage{float}
\usepackage{amsmath,amsfonts,amssymb}
\usepackage{booktabs}
\usepackage{subcaption}
\usepackage[skip=2pt]{caption}
\usepackage[update,prepend]{epstopdf}
\usepackage[lined,algonl,ruled]{algorithm2e}
\usepackage{balance}
\usepackage{acronym}
\usepackage{color}
\usepackage{multirow}
\usepackage{multicol}
\usepackage{hyperref}

\acrodef{MoE}{Mixture of Experts}
\acrodef{SOTA}{State of The Art}
\acrodef{ATF}{Acoustic Transfer Function}
\acrodef{RIR}{Room Impulse Response}
\acrodef{SNR}{Signal-to-Noise Ratio}
\acrodef{MULCAT}{Multiply-and-Concatenate}
\acrodef{CNN}{Convolutional Neural Network}
\acrodef{SI-SNR}{Scale-Invariant Signal-to-Noise Ratio}
\acrodef{uPIT}{utterance level Permutation Invariant Training}
\acrodef{STFT}{Short-Time Fourier Transform}
\acrodef{WSJ}{Wall Street Journal}

\usepackage{bm}
\usepackage{amssymb,amsmath,graphicx,bbm,color,booktabs}
\usepackage{amsopn}

\newcommand{\ve}{\bm{e}}

\newcommand{\vech}{\bm{h}}

\newcommand{\vn}{\bm{n}}               
\newcommand{\vo}{\bm{o}}

\newcommand{\vs}{\bm{s}}       \newcommand{\vsh}{\hat{\bm{s}}}        
               
\newcommand{\vu}{\bm{u}}               
\newcommand{\vv}{\bm{v}}               
               
\newcommand{\vx}{\bm{x}}               
               
\newcommand{\vz}{\bm{z}}               

\newcommand{\R}{\mathbb{R}}

\renewcommand{\S}{\mathcal{S}}

\renewcommand{\eqref}[1]{Eq.~(\ref{#1})}

\title{single channel voice separation for unknown number\\ of speakers under reverberant and noisy settings}
\name{Shlomo E. Chazan$^1*$, Lior Wolf$^1$, Eliya Nachmani$^{1,2}$, Yossi Adi$^1$\thanks{Samples: \url{https://shlomke.github.io/Samples/ICASSP_2021}. *Work done while Shlomo was an Intern at Facebook AI Research}}
 \address{$^1$Facebook AI Research, 
        $^2$Tel Aviv University \vspace{-0.5cm}}

\begin{document}
\maketitle

\begin{abstract}
We present a unified network for voice separation of an unknown number of speakers. The proposed approach is composed of several separation heads optimized together with a speaker classification branch. The separation is carried out in the time domain, together with parameter sharing between all separation heads. The classification branch estimates the number of speakers while each head is specialized in separating a different number of speakers. We evaluate the proposed model under both clean and noisy reverberant settings. Results suggest that the proposed approach is superior to the baseline model by a significant margin. Additionally, we present a new noisy and reverberant dataset of up to five different speakers speaking simultaneously. 
\end{abstract}


\vspace{-0.2cm}
\section{Introduction}
\vspace{-0.2cm}
\label{sec:intro}

In real-world acoustic environments, a speech signal is frequently corrupted by a noisy environment, room conditions, multi-talker setup, etc. The ability to separate a single voice from multiple conversations is crucial for any speech processing system designed to perform under such conditions. Over the years, many attempts have been made to tackle this separation problem considering single microphone~\cite{hershey2016deep, yu2017permutation}, multiple microphones~\cite{gannot2017survey,makino2018audio}, supervised and unsupervised learning~\cite{hyvarinen2000independent,seetharaman2018bootstrapping}. 

In this work, we focus on fully supervised voice separation using a single microphone, which has seen a great leap in performance following the recent success of deep learning models considering both frequency domain~\cite{hershey2016deep, yu2017permutation, upit, chen2017deep, wang2018alternative, wang2019deep}, and time-domain~\cite{luo2019conv, stoller2018wave, venkataramani2018end, luo2020dual, zhang2020furcanext, zeghidour2020wavesplit} modeling.



Despite its success, most prior work assumes the number of speakers in the mixture to be known a-priori. Recently, several studies proposed various methods to tackle this problem. The authors of~\cite{kinoshita2018listening,takahashi2019recursive,shi2020sequence} suggest to separate \emph{one speaker at a time} using a recursive solution. This requires $C$ sequential forward passes to separate $C$ sources and it is not clear when to stop the separating process. The authors of~\cite{von2020multi} proposed a similar \emph{one speaker at a time} solution however they were mainly interested in automatic speech recognition as the final downstream task. Another line of prior work, optimize the network to output the maximum number of speakers regardless of the actual number of speakers present in the input mixture. At test time, the number of speakers is determined by detecting the number of silent channels~\cite{upit, luo2018speaker}. Although this method is shown to perform well, it was evaluated only under an anechoic setup while considering up to three speakers. 

\begin{figure}[t!]  
	\centering
	\includegraphics[scale=0.45]{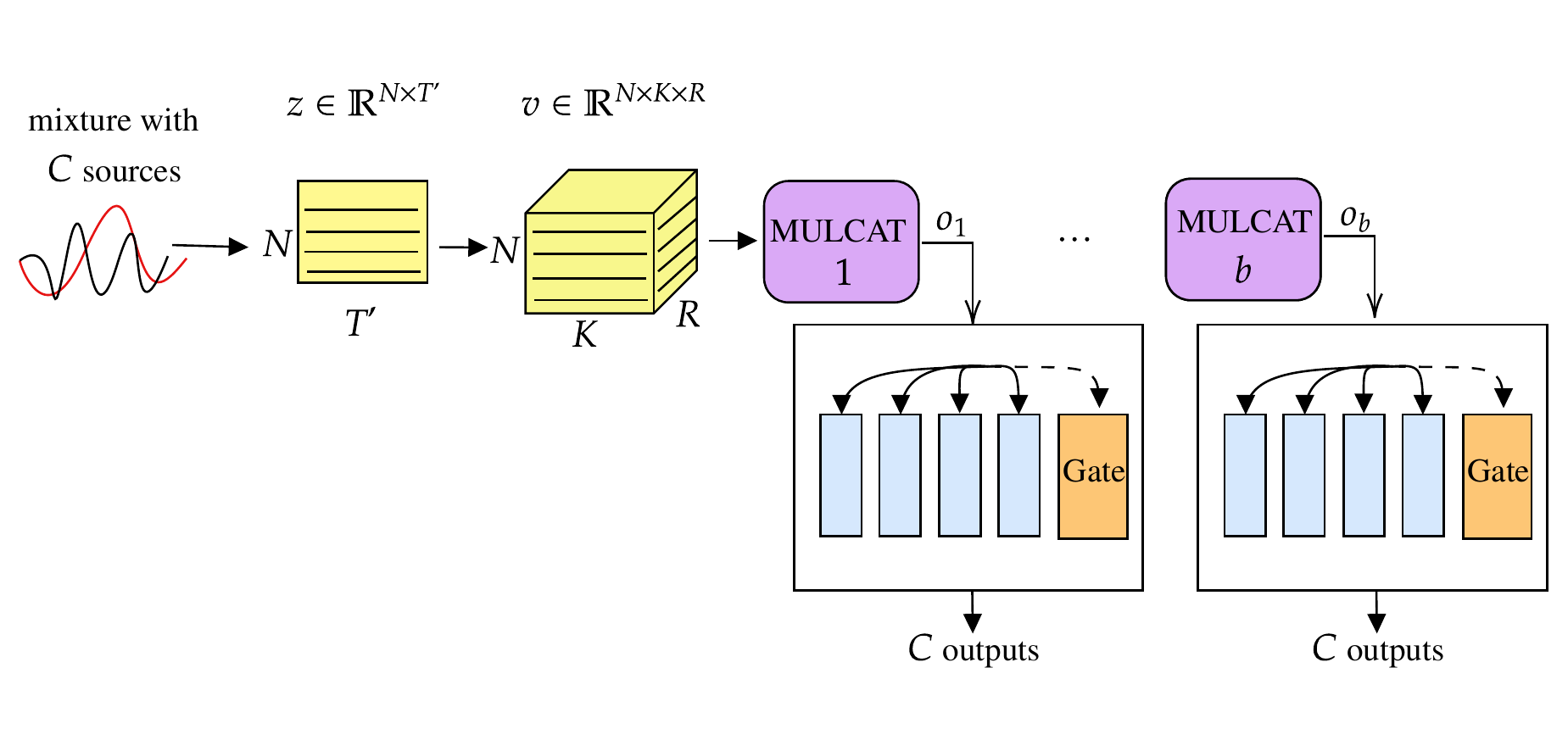}
	\caption{The architecture of the proposed network. The feature extraction constructed with 1D convolutions and chunking. Then $b$ units are applied using the same separation heads to produce output after each block. \vspace{-0.3cm}}
	\label{fig:arch}
\end{figure}

The most relevant prior work to ours is~\cite{nachmani2020voice}. In this study, the authors suggested training several models, each for separating a different number of speakers. A model selection heuristic is applied on top of the obtained models predictions to detect non-activated channels (noise / silence). Despite its success, it has two main drawbacks. First, several different models were trained separately, hence at test time the input mix is propagating throughout each separately. This makes inference costly in terms of memory and computing power. Additionally, training each model separately does not benefit from shared representations, e.g., the representation learned while separating two speakers can be beneficial for separating four speakers. Second, under the unknown number of speakers setting only anechoic setup was considered. While \cite{nachmani2020voice} reported results on WHAMR! dataset~\cite{whamr}, which contains noisy reverberant mixtures, this dataset consists of mixtures of two sources only.

In this study, we propose a unified approach to separate up to five different speakers simultaneously speaking using several separation heads together with shared representations. Our model is working directly on the raw waveform and was evaluated under both anechoic and noisy reverberant environments. The proposed model obtains superior performance over the baseline methods, especially when considering the number of speakers in the mixture to be unknown. We additionally release the scripts used to generate the proposed noisy reverberant datasets.

\vspace{-0.3cm}
\section{Problem Setting}
\vspace{-0.2cm}
\label{sec:probelm}

\subsection{Anechoic room}
Consider a single microphone, recording a mixture of $C$ different sources $\vs^j \in \mathbb{R}^T$, where $j \in \left[1,\ldots,C\right]$ in an anechoic enclosure where the source length, $T$ can vary. The mixed signal is therefore: $\vx=\sum_{j=1}^{C} \alpha^j\cdot \vs^j$,
%
where $\alpha^j$ is the scaling factor of the $j$-th source. Although this model is commonly used to demonstrate separation abilities, anechoic noiseless environments are hard to find in the real world. 

\subsection{Noisy reverberant room}
To simulate a more real-world setting an \ac{ATF} which relate the sources and the microphones is considered together with additive noise as follows: $\vx=\sum_{j=1}^{C} \alpha^j\cdot \vs^j*\vech^j + \vn,$
%
where $\vech^j$ is the \ac{ATF} of the $j$-th source to the microphone, and $\vn$ is a non stationary additive noise in an unknown \ac{SNR}. 

Under both cases, we focus on the fully supervised setting, in which we are provided with a training set $\S = \{\vx_i, (\vs_{i}^1, \cdots, \vs_{i}^C)\}_{i=1}^m$, and our goal is learn a model that given an unseen mixture $\vx$, outputs $C$ separate channels, $\vsh$, that maximize the \ac{SI-SNR} to the ground truth signals when considering reordering of the output channels $(\vsh^{\pi(1)}, \cdots, \vsh^{\pi(C)})$ for the optimal permutation $\pi$.

\vspace{-0.3cm}
\section{Model}
\vspace{-0.2cm}
\label{sec:model}
We propose to jointly separate a varying number of sources using a single model with several separation heads and shared representations. The proposed architecture is depicted in Fig.~\ref{fig:arch}.

Following the architecture proposed in~\cite{luo2020dual}, the mixed signal is first encoded using a stack of $N$ 1D convolution with a kernel size of $L$ and stride of $L/2$ followed by ReLU function. The 2D tensor output of the encoder is given by $\vz\in \mathbb{R}^{N\times T'}$, where $T'=(2T/L)-1$. Next, $\vz$ is going through a chunking process. It is first divided into $R$ overlapping chunks with chunk size of $K$ and step size of $P$, denoted as $\vu_r \in \R^{N\times K}$, where $r \in [1, \cdots , R ]$. Then the 2D chunks are concatenated into a 3D embedded tensor $\vv=\left[\vu_1,\ldots,\vu_R\right] \in \mathbb{R}^{N\times K \times R}$. Next, a series of $b$ \ac{MULCAT} blocks, as proposed in~\cite{nachmani2020voice}, are employed to model the intra-chunk and inter-chunk dependencies. 

We separate the mixture using several separation heads after each block $l\in \{ 1,\ldots,b \}$ and output $\vo_l$. The separation heads architecture is containing four experts alongside a gate. The $n$-th expert' expertise is to separate different number of speakers $C_n$, where $n\in \{ 1,\ldots,4 \}$ and  $C_n \in \{ 2,3,4,5 \}$, respectively.  Note, all the experts and the gate share the same input $\vo_l$. Each expert is comprised of a PReLU non-linearity with parameters initialized at 0.25, followed by $1\times 1$ convolution with $C_n\cdot R$ kernels. The resulting tensor with a size of $N\times K\times C_n\cdot R$ is then divided into $C_n$ tensors with size $N\times K\times R$, which are finally transformed to $C_n$ waveforms samples by applying an overlap-and-add operation to the $R$ chunks. The overlap between two successive frames is $L/2$.

The gating network is implemented as \ac{CNN} using four convolution layers with   $64,32,16,8$ channels, respectively, followed by two fully connected layers. Each convolutional layer has a kernel size of 3 followed by PReLU and max-pooling with kernel size 2. The first fully connected layers have 100 PReLU neurons while the last layer outputs a distribution over the number of speakers. Unlike~\cite{nachmani2020voice}, we do not use any speaker identification loss. Note, that the same separation heads are applied after each block.


\vspace{-0.2cm}
\paragraph*{Training objective}
We optimize several loss functions to further improve models performance, where the main objective of each of the experts is the \ac{SI-SNR}, 

\begin{equation}
    \text{SI-SNR}(\vs^j,\vsh^j)=10\log_{10}\frac{||\Tilde{\vs}^j||^2}{\|\Tilde{\ve}^j\|^2}, 
\end{equation} 
where $\Tilde{\vs}^j=\frac{\langle \vs^j,\vsh^j\rangle \vs^j}{||{\vs}^j||^2}$ and $\Tilde{\ve}^j=\vsh^j-\Tilde{\vs}^j$. 

To tackle the permutation invariant problem we use the \ac{uPIT} loss, as proposed in~\cite{upit}:
\begin{equation}
    L_{\text{uPIT}}(\vs,\vsh)=-\max_{\pi\in \Pi_{C_n}} \frac{1}{C_n}\sum_{j=1}^{C_n}\text{SI-SNR}(\vs^j,\vsh^{\pi(j)}),
    \label{eq:si_snr_pit}
\end{equation}
where $\Pi_{C_n}$ is the set of all possible permutations of $1,\ldots,C_n$. We denote the optimal permutation $\pi_{o}$.

Next, we include a frequency domain loss function. Similarly to~\cite{parallel_wavgan,yamamoto2019probability}, we define the STFT loss to be the sum of the \emph{spectral convergence (sc)} loss and the \emph{magnitude} loss as follows,
\begin{equation}
    \label{eq:mrstft}
    \begin{aligned}
    &L_{\text{stft}}=\sum_{j=1}^{C_n} L_{sc}(\vs^j, \vsh^{\pi_{o}(j)}) + L_{mag}(\vs^j, \vsh^{\pi_{o}(j)}), \\
    &L_{\text{sc}}(\vs^j, \vsh^{\pi_{o}(j)}) = \frac{\| |\text{STFT}(\vs^j)| - |\text{STFT}(\vsh^{\pi_{o}(j)})|\|_F}{\||\text{STFT}(\vs^j)| \|_F},\\
    &L_{\text{mag}}(\vs^j, \vsh^{\pi_{o}(j)}) = \frac{1}{T}\| \log|\text{STFT}(\vs^j)| - \log|\text{STFT}(\vsh^{\pi_{o}(j)})|\|_1,
    \end{aligned}
\end{equation}
where $\| \cdot \|_F$ and $\| \cdot \|_1$ are the Frobenius the $L_1$ norms respectively. We define the multi-resolution STFT loss to be the sum of all STFT loss functions using different STFT parameters. We apply the STFT loss using different resolution with number of FFT bins $\in \{512, 1024, 2048\}$, hop sizes $\in \{50, 120,240 \}$, and lastly window lengths $\in \{240, 600, 1200\}$.

Lastly, we included a cyclic reconstruction L2 loss between the sum of the input mixture to the sum of the estimated sources. Defined as: $L_{\text{rec}}=\| \sum_{j=1}^{C_n}\vsh^j - \vx \|^2$. Notice, in the case of noisy and reverberant setup, we replace $\vx$ by the sum of all clean input sources.  

Overall, we minimize the following objective function,
\begin{equation}
    \label{eq:final_loss}
    L=L_{\text{uPIT}}+\lambda_{stft}\cdot L_{\text{stft}}+\lambda_{rec}\cdot L_{\text{rec}}+ \lambda_{\text{gate}}\cdot L_{g}, 
\end{equation}
where $L_{g}$ is the categorical cross-entropy loss used to optimize the gate branch. Note, the gate is constantly training regardless of the amount of sources. We calibrated all $\lambda$ values on the validation set, and set $\lambda_{\text{rec}}=\lambda_{\text{gate}}=1$, and  $\lambda_{\text{stft}}=0.5$. 

At the training phase, the number of speakers, $C_n$ is randomly chosen in each mini-batch. Therefore, only the corresponding expert is trained at every mini-batch. During inference, the outputs of the expert with the highest probability are used.

\vspace{-0.1cm}
\paragraph*{Evaluation method}
While evaluating a separation model for a known the number of speakers is straightforward and can be done by using \ac{SI-SNR} directly, it is unclear how to evaluate a separation model with an unknown number of speakers, since the predicted and target number of speakers can vary. 

To mitigate that we follow the method proposed by~\cite{nachmani2020voice}. Three cases are considered: i) the predicted and target number of speakers are the same, in this case, we simply compute the SI-SNR; ii) the predicted number of speakers is larger than the target number of speakers, here we compute the correlation between each predicted and target channels, and pick the $C$ predicted channels with the highest correlation; iii) the predicted number of speakers is smaller than the target number of speakers. Here we also compute the correlation between the predicted and target channels, but then we duplicate the best-correlated signals to reach $C$ number of channels. 

The last case can be considered as a penalty for the model since the separation will always be flawed. In the second case, the model may produce a good separation despite predicting the wrong number of speakers.

\begin{table}[t!]
\caption{Noisy reverberant data specification.}
\label{table:noisy_data}
\centering
\resizebox{0.6\columnwidth}{!}{
\begin{tabular}{@{}lll@{}}
\toprule
                        & x        & \emph{U}{[}4,7{]}                                  \\
Room (m)                & y        & \emph{U}{[}4,7{]}                                  \\
                        & z        & 2.5                                         \\ \midrule
T\_60 (sec)             &          & \emph{U}{[}0.16, 0.36{]}                           \\ \midrule
                        & x        & $\frac{x_{\text{Room}}}{2}$+\emph{U}{[}-0.2,0.2{]} \\
Mic. Pos. (m)           & y        & $\frac{y_{\text{Room}}}{2}$+\emph{U}{[}-0.2,0.2{]} \\
                        & z        & 1.5                                         \\ \midrule
\# of speakers          &          & \{2/3/4/5\}                                 \\ \midrule
Sources Pos. ($^\circ$) & $\theta$ & \emph{U}{[}0,180{]}                                \\ \midrule
Sources Distance (m)                   &          & 1.5+\emph{U}{[}-0.2,0.2{]}                         \\ \midrule
SNR                     & dB       & \emph{U}{[}0, 15{]}                                 \\ \bottomrule
\end{tabular}}
\vspace{-0.3cm}
\end{table}

\vspace{-0.2cm}
\section{Dataset}
\label{sec:noisy_dataset}
\vspace{-0.2cm}
Under both clean and noisy settings, we assume all signals were sampled at $8$~kHz. We set 20,000 examples for training, 5,000 samples for validation, and 3,000 samples for testing. We consider the anechoic signals as target supervision, thus under the noisy reverberant setting, we optimize the model to jointly do separation, denoising, and dereverberation. 

\paragraph*{Clean dataset}
For the clean dataset, we use the wsj0-2mix and wsj0-3mix mixtures as suggested in~\cite{hershey2016deep}, while for wsj0-4mix and wsj0-5mix we follow the same mixture recipes as suggested in~\cite{nachmani2020voice}.

\paragraph*{Noisy reverberant dataset}
As for the noisy reverberant settings, we generate datasets for separating up to five different sources. The setup of the dataset is presented in Table~\ref{table:noisy_data}. We synthetically generate noisy reverberant mixtures to mimic real-world recordings. The clean signals were taken from the WSJ0 corpus~\cite{WSJ0} and noise signals from the WHAM! noise dataset~\cite{WHAM}. 

For each mixture, we randomly selected room dimensions, microphone positions, and different positions for the sources, as shown in Table~\ref{table:noisy_data}. We generated a \ac{RIR}  using the rir\_generator tool~\cite{rir_generator} for every speaker in the mixture,  which was then convolved with the clean signal. The reverberant signals were then summed up together with an additive noise using random \ac{SNR}. 
\vspace{-0.2cm}
\section{Experimental Results}
\vspace{-0.2cm}
\label{ssec:experiments}
We start by evaluating the proposed model while we assume the number of speakers in the mixture is known a-priori. Next, we move into comparing our system to several automatic-selection methods while the number of speakers in the recording is unknown. We conclude this section by analyzing the performance of the speaker classification branch. All results are reported for both clean and noisy reverberant environments. For the separation results, we report the \ac{SI-SNR} improvement over the mixture, denoted as SI-SNRi. 

\vspace{-0.2cm}
\subsection{Known number of speakers}
\label{subssec:known}
We compared the proposed method to ConvTasNet~\cite{luo2019conv}, Dual-Path RNN (DPRNN)~\cite{luo2020dual}, and Gated model~\cite{nachmani2020voice}, for the case of a known number of speakers. The baseline methods were trained with a different model separating each number of speakers between two and five. We optimized all baseline models using the published code by the method's authors. All models were optimized until no loss improvement was observed on the validation set for five epochs using Adam optimizer with a learning rate of $3\times10^{-4}$ and a batch size of 2.

Table~\ref{tab:known} presents the separation results. The proposed method is superior to the baseline methods by a significant margin, with one exception of two speakers in an anechoic room. These results suggest that using shared representation together with classifying the number of speaker in the mixture are beneficial specifically when considering more than two speakers or a noisy environment. 

Notice, the noisy dataset is significantly more challenging than the clean dataset since the models are required to not only separate the sources but also reduce their reverberation and additive noise. Therefore all models suffer a degradation in performance compared to the clean dataset. 
\begin{table}[!t]
    \centering
    \caption{Performance of various models as a function of the number of speakers under the clean and noisy reverberant setups. In the following results, we assume the number of speakers in the mixture is known a-priori. All results are reported in SI-SNRi.}
    \label{tab:known}
    \resizebox{\columnwidth}{!}{%
    \begin{tabular}{l | cccc | cccc}
    \toprule
    Model& 2spk&	3spk&	4spk&	5spk & 2spk&	3spk&	4spk&	5spk\\
    \midrule
    \multicolumn{1}{c|}{} & \multicolumn{4}{c|}{Clean} & \multicolumn{4}{c}{Noisy-reverberant} \\
    \midrule
    ConvTasNet~\cite{luo2019conv}	& 15.33				&	12.71  			&	8.52		         &	7.04 			& 8.97 			 &	7.46 			&   6.31			& 	5.53  \\
    DPRNN~\cite{luo2020dual}		& 18.81				&	14.68 			&	10.39	             &	8.72				& 10.24		 	 &	8.34			&	6.92			&	5.89  \\
    Gated~\cite{nachmani2020voice}		& \textbf{20.12}   	&	 16.85				&	12.88				 &	10.56 		    & 10.66			 &	8.93			&	7.42			&	6.35 \\
    Ours		& 19.43 				&	\textbf{17.26} 	&	\textbf{13.93} 	 &	\textbf{11.77}  & \textbf{11.48} & \textbf{10.73} 	& \textbf{9.48} 	& \textbf{8.49}\\
    \bottomrule
    \end{tabular}}
\end{table}

\vspace{-0.2cm}
\subsection{Unknown number of speakers}
\label{subssec:unknown}
Next, we consider the case of an unknown number of speakers. We compared the proposed method to several automatic selection algorithms for the number of speakers in the recording. Specifically, we compared our model to i)~\cite{nachmani2020voice} which trained a separate model to separate a different number of speakers, denoted as Ensemble; ii)~\cite{upit, luo2018speaker} which trains one model to separate the maximum number of speakers, denoted as MaxOut. We optimized the MaxOut method with and without speaker classification loss. Notice, both methods use a silent detection algorithm on top of the model's output to produce the final separation. In contrast, our work uses a speaker classification branch, we use its output to determine the number of speakers in the mixture. 

For a fair comparison, all separation models are based on Gated~\cite{nachmani2020voice}, where we only change the selection algorithm. Results presented in Table~\ref{tab:unknown}. The proposed method is superior to the baseline methods under both clean and noisy scenarios. Notice, sharing internal representation yields in a better separation performance, while including several separation heads instead of the MaxOut method further improves the results, specifically under noisy environments. Interestingly, including the classification branch did not improve performance for the MaxOut method.

\begin{table}[t!]
    \centering
    \caption{A comparison of several automatic selection algorithms for speaker separation while considering the number of speakers in the mixture to be unknown. All results are reported in SI-SNRi.}
    \label{tab:unknown}
    \resizebox{\columnwidth}{!}{%
    \begin{tabular}{l | cccc | cccc}
    \toprule
    Model& 2spk&	3spk&	4spk&	5spk & 2spk&	3spk&	4spk&	5spk\\
    \midrule
    \multicolumn{1}{l|}{} & \multicolumn{4}{c|}{Clean} & \multicolumn{4}{c}{Noisy-reverberant} \\
    \midrule
    Ensemble~(\cite{nachmani2020voice})			& 18.63    			&	14.62   		&	11.48    		&	10.37  			&	10.24			&	8.59 			&	7.07				&	6.21\\
    MaxOut w/o Cls.~(\cite{upit, luo2018speaker})			& 19.29		   			&	16.8   			&	13.34				&	11.31  			&	10.59			&	9.41 			&	7.92				&	7.5\\
    \hline
    MaxOut w/ Cls.~(\cite{upit, luo2018speaker})			& 19.11   			&	16.71   			&	13.35				&	11.29  			&	10.58			&	9.39 			&	7.97				&	7.51\\
    Ours										& \bf{19.41}		    &	\bf{17.05}  	&	\bf{13.91}  	&	\bf{11.71} 		&	\bf{11.45}	&	\bf{10.6}	&	\bf{9.36}		&	\bf{8.31}\\
    \bottomrule
    \end{tabular}}
\end{table}

\begin{figure}[t!]
\begin{subfigure}{0.239\textwidth}
  \centering
  \includegraphics[width=.8\linewidth]{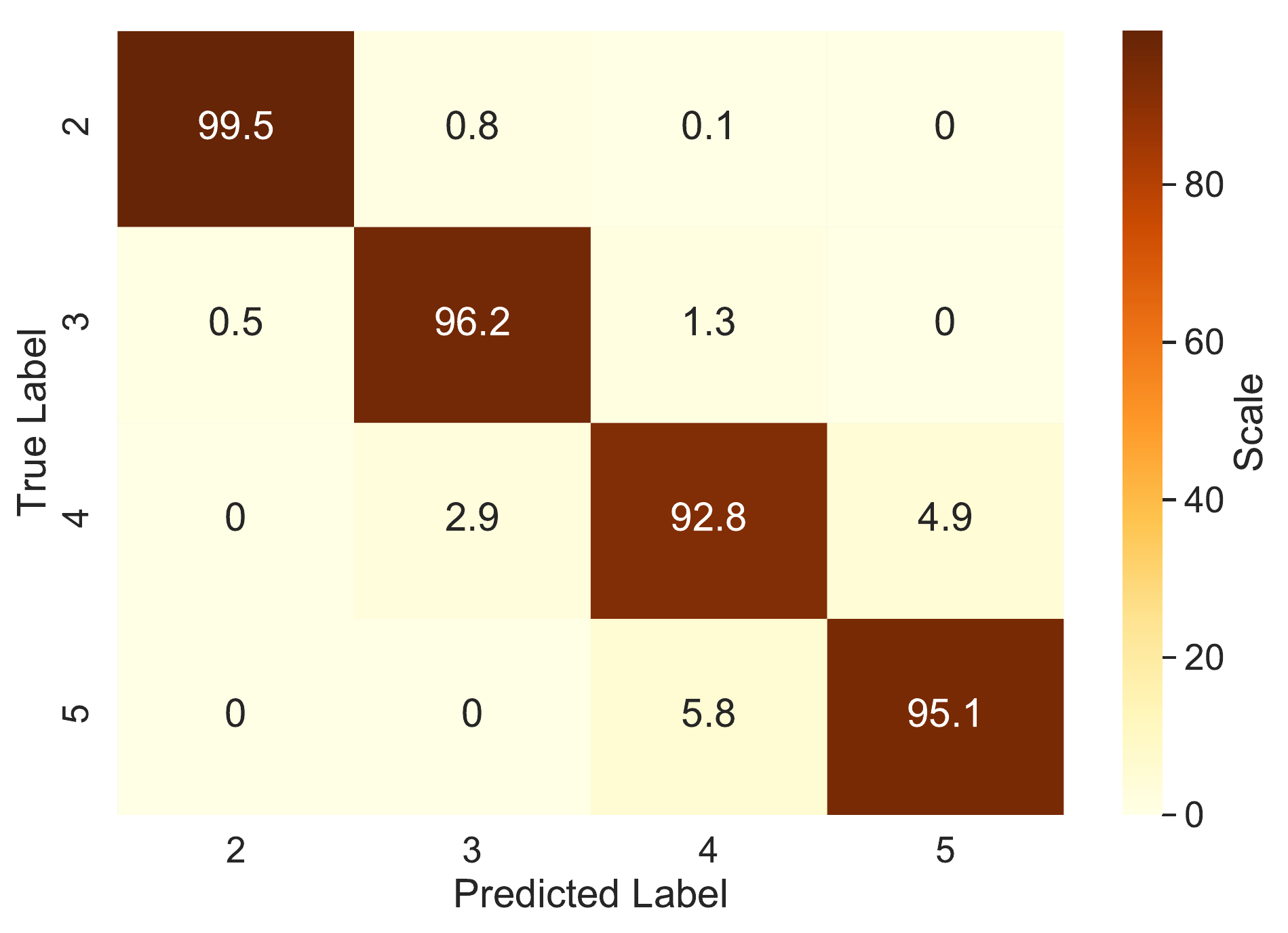}  
  \caption{}
  \label{fig:conf_clean_ours}
\end{subfigure}
\begin{subfigure}{0.239\textwidth}
  \centering
  \includegraphics[width=.8\linewidth]{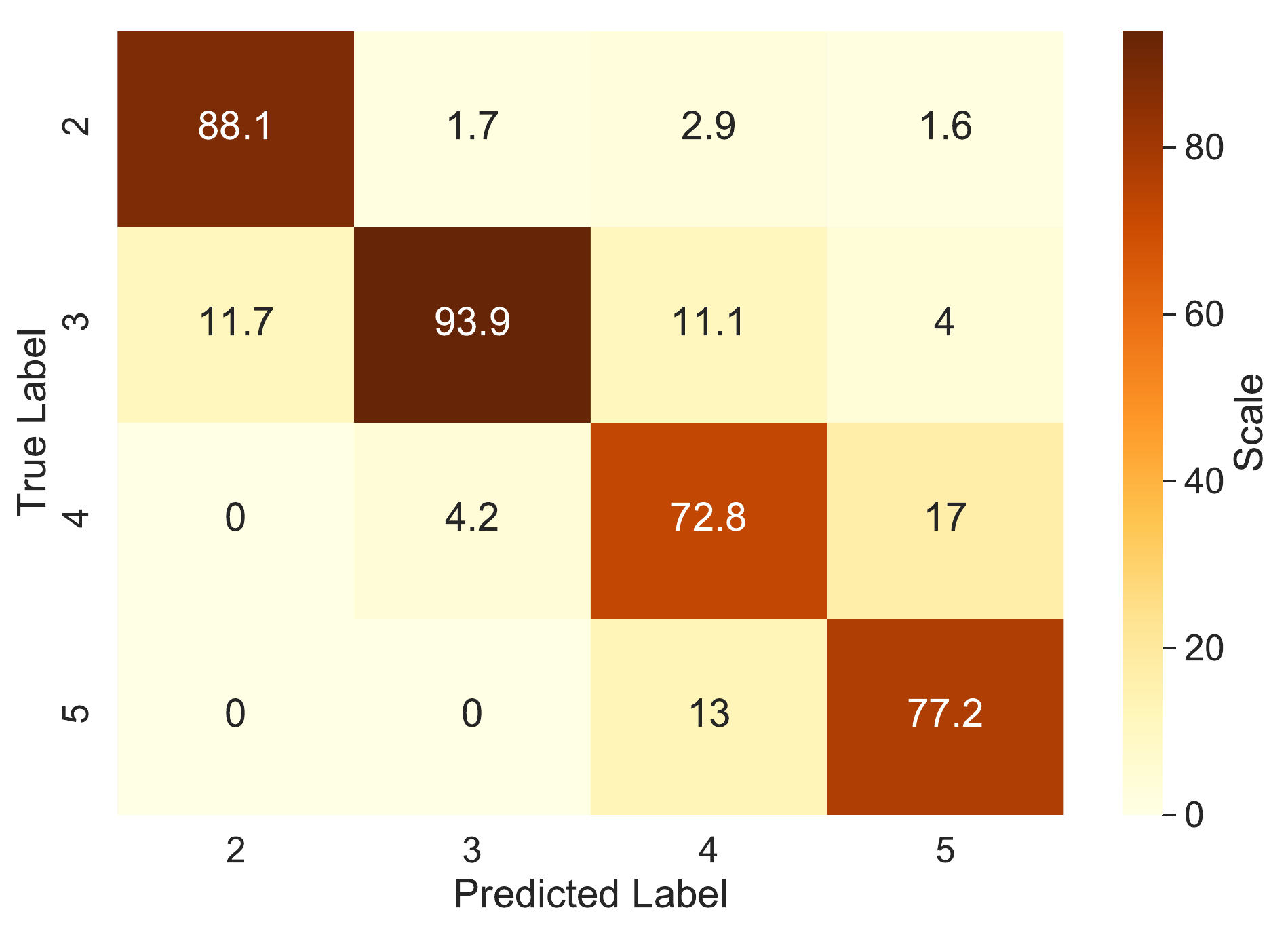}  
  \caption{}
  \label{fig:conf_clean_icml}
\end{subfigure}
\\
\begin{subfigure}{0.239\textwidth}
  \centering
  \includegraphics[width=.8\linewidth]{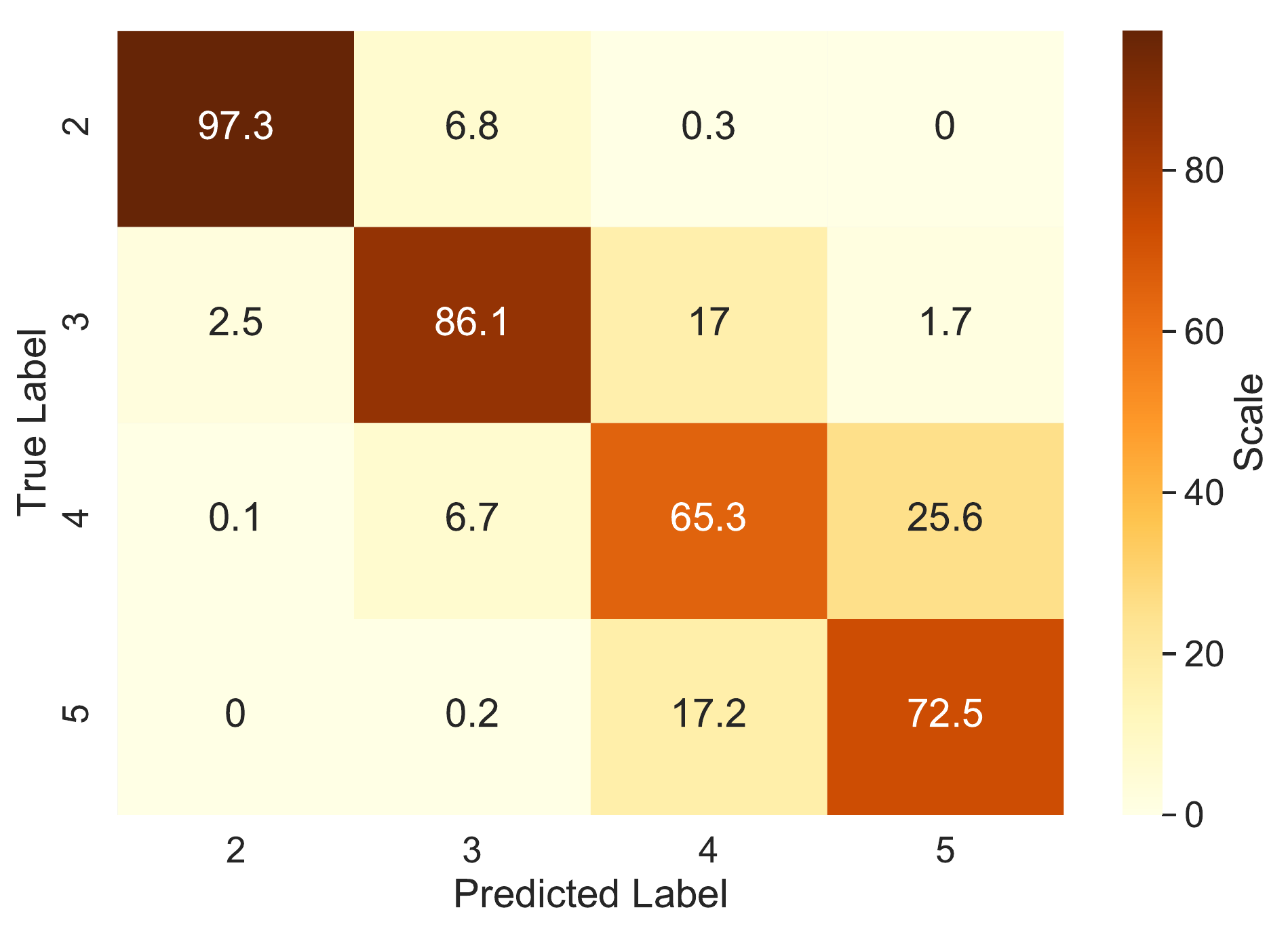}  
  \caption{}
  \label{fig:conf_noisy_ours}
\end{subfigure}
\begin{subfigure}{0.239\textwidth}
  \centering
  \includegraphics[width=.8\linewidth]{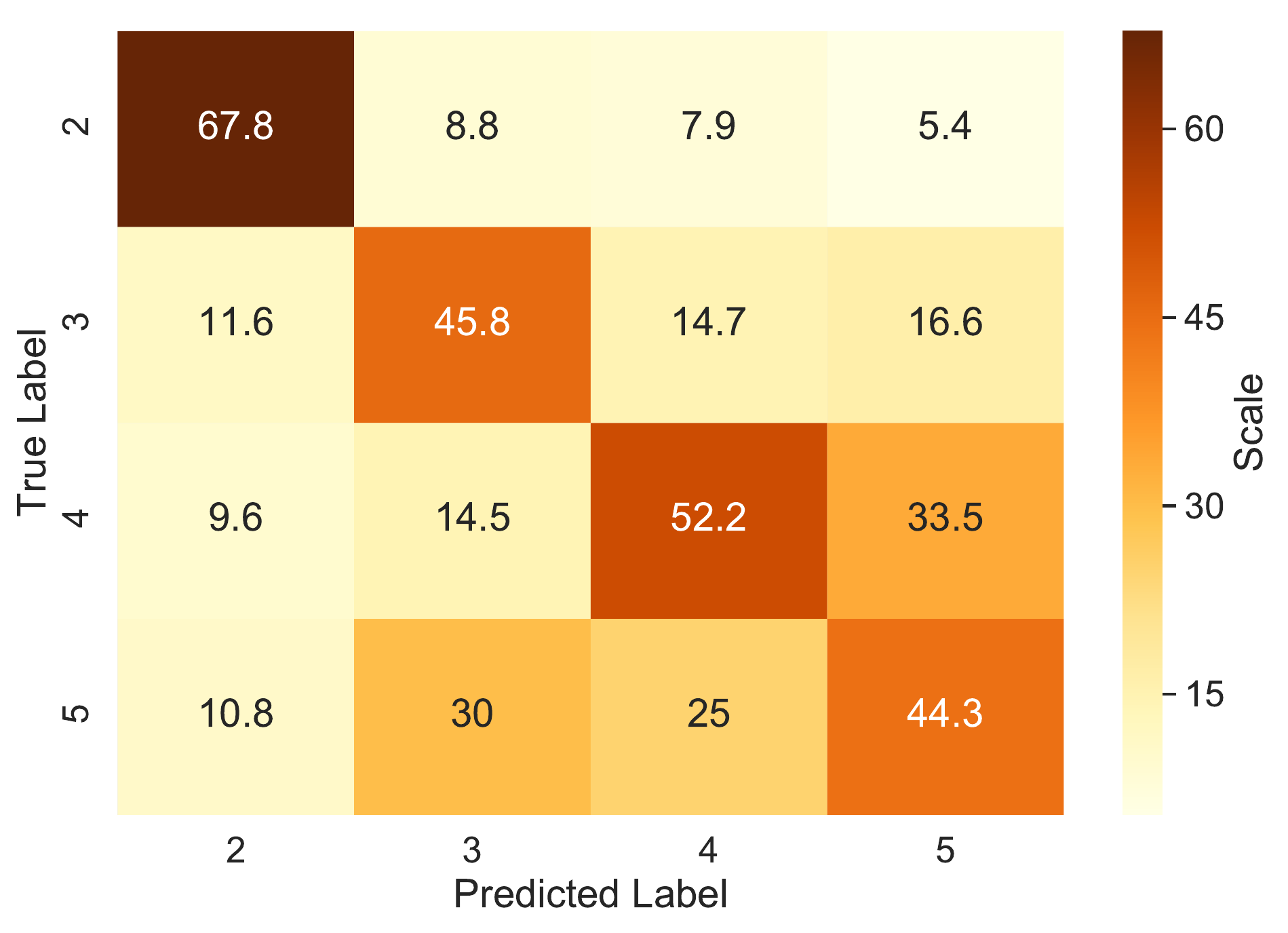}  
  \caption{}
  \label{fig:conf_noisy_icml}
\end{subfigure}
\caption{Confusion matrix for model selection results using clean and noisy datasets. Results are reported for both the proposed model (Fig.~\ref{fig:conf_clean_ours} (clean) and Fig.~\ref{fig:conf_noisy_ours} (noisy)) and the MaxOut model using silent detection method as proposed in~\cite{nachmani2020voice} (Fig.~\ref{fig:conf_clean_icml} (clean) and Fig.~\ref{fig:conf_noisy_icml} (noisy)). Acc. is presented inside each cell in the matrix.
\vspace{-0.35cm}}
\label{fig:conf}
\end{figure}

Lastly, we report the classification results obtained by our model and compared them to the silent detection algorithm as in~\cite{nachmani2020voice}. The results are depicted in Fig.~\ref{fig:conf}. Including a dedicated branch for speaker separation evidently provides a boost in classification performance, especially in noisy reverberant environments. As a side-note: we also experimented with optimizing the classification model using spectral feature rather than joint optimization with the separation heads. This, however, provided inferior performance.

It is worth mentioning that although SI-SNRi results are superior to the baseline methods while listening to the separations there still much room for improvement, especially when considering the mixtures with four or five speakers under noisy reverberant environments.
Nevertheless, these separations can still be used as prior statistics for next-phase multi-channel speech processing.

\vspace{-0.3cm}
\section{Conclusions}
\vspace{-0.2cm}
\label{sec:conclusions}
Single-channel source separation is a challenging task, especially when considering a large or unknown number of speakers in noisy reverberant environments.  In this work, we introduce a neural net model that handles the uncertainty regarding the number of speakers under real-world conditions. The success of our work under practical settings stems from the use of a shared representation with a multi-task loss function. Empirical results suggest the proposed method is superior to the baseline models both in terms of separation and classifying the number of speakers in the mixture.

\bibliographystyle{IEEEbib}
\bibliography{refs, icml}

\begin{thebibliography}{10}

\bibitem{hershey2016deep}
John~R Hershey et~al.,
\newblock ``Deep clustering: Discriminative embeddings for segmentation and
  separation,''
\newblock in {\em ICASSP}, 2016.

\bibitem{yu2017permutation}
Dong Yu et~al.,
\newblock ``Permutation invariant training of deep models for
  speaker-independent multi-talker speech separation,''
\newblock in {\em ICASSP}, 2017.

\bibitem{gannot2017survey}
S.~Gannot et~al.,
\newblock ``A consolidated perspective on multimicrophone speech enhancement
  and source separation,''
\newblock {\em IEEE/ACM Transactions on Audio, Speech, and Language
  Processing}, pp. 692--730, Apr. 2017.

\bibitem{makino2018audio}
Shoji Makino,
\newblock {\em Audio Source Separation},
\newblock Springer, 2018.

\bibitem{hyvarinen2000independent}
Aapo Hyv{\"a}rinen and Erkki Oja,
\newblock ``Independent component analysis: algorithms and applications,''
\newblock {\em Neural networks}, vol. 13, no. 4-5, pp. 411--430, 2000.

\bibitem{seetharaman2018bootstrapping}
Prem Seetharaman et~al.,
\newblock ``Bootstrapping single-channel source separation via unsupervised
  spatial clustering on stereo mixtures,''
\newblock {\em arXiv preprint arXiv:1811.02130}, 2018.

\bibitem{upit}
M.~{Kolbæk} et~al.,
\newblock ``Multitalker speech separation with utterance-level permutation
  invariant training of deep recurrent neural networks,''
\newblock {\em IEEE/ACM Transactions on Audio, Speech, and Language
  Processing}, 2017.

\bibitem{chen2017deep}
Zhuo Chen, Yi~Luo, and Nima Mesgarani,
\newblock ``Deep attractor network for single-microphone speaker separation,''
\newblock in {\em ICASSP}, 2017.

\bibitem{wang2018alternative}
Zhong-Qiu Wang et~al.,
\newblock ``Alternative objective functions for deep clustering,''
\newblock in {\em ICASSP}, 2018.

\bibitem{wang2019deep}
Zhong-Qiu Wang, Ke~Tan, and DeLiang Wang,
\newblock ``Deep learning based phase reconstruction for speaker separation: A
  trigonometric perspective,''
\newblock in {\em ICASSP}, 2019.

\bibitem{luo2019conv}
Yi~Luo and Nima Mesgarani,
\newblock ``Conv-tasnet: Surpassing ideal time--frequency magnitude masking for
  speech separation,''
\newblock {\em IEEE/ACM transactions on audio, speech, and language
  processing}, pp. 1256--1266, 2019.

\bibitem{stoller2018wave}
Daniel Stoller, Sebastian Ewert, and Simon Dixon,
\newblock ``Wave-u-net: A multi-scale neural network for end-to-end audio
  source separation,''
\newblock {\em arXiv preprint arXiv:1806.03185}, 2018.

\bibitem{venkataramani2018end}
Shrikant Venkataramani and Paris Smaragdis,
\newblock ``End-to-end source separation with adaptive front-ends,''
\newblock {\em CoRR}, vol. abs/1705.02514, 2017.

\bibitem{luo2020dual}
Yi~Luo, Zhuo Chen, and Takuya Yoshioka,
\newblock ``Dual-path rnn: efficient long sequence modeling for time-domain
  single-channel speech separation,''
\newblock in {\em ICASSP}, 2020.

\bibitem{zhang2020furcanext}
Liwen Zhang et~al.,
\newblock ``Furcanext: End-to-end monaural speech separation with dynamic gated
  dilated temporal convolutional networks,''
\newblock in {\em International Conference on Multimedia Modeling}, 2020.

\bibitem{zeghidour2020wavesplit}
Neil Zeghidour and David Grangier,
\newblock ``Wavesplit: End-to-end speech separation by speaker clustering,''
\newblock {\em arXiv preprint arXiv:2002.08933}, 2020.

\bibitem{kinoshita2018listening}
Keisuke Kinoshita et~al.,
\newblock ``Listening to each speaker one by one with recurrent selective
  hearing networks,''
\newblock in {\em ICASSP}, 2018.

\bibitem{takahashi2019recursive}
Naoya Takahashi et~al.,
\newblock ``Recursive speech separation for unknown number of speakers,''
\newblock {\em arXiv preprint arXiv:1904.03065}, 2019.

\bibitem{shi2020sequence}
Jing Shi et~al.,
\newblock ``Sequence to multi-sequence learning via conditional chain mapping
  for mixture signals,''
\newblock {\em arXiv preprint arXiv:2006.14150}, 2020.

\bibitem{von2020multi}
Thilo von Neumann et~al.,
\newblock ``Multi-talker asr for an unknown number of sources: Joint training
  of source counting, separation and asr,''
\newblock {\em arXiv preprint arXiv:2006.02786}, 2020.

\bibitem{luo2018speaker}
Yi~Luo, Zhuo Chen, and Nima Mesgarani,
\newblock ``Speaker-independent speech separation with deep attractor
  network,''
\newblock {\em IEEE/ACM Transactions on Audio, Speech, and Language
  Processing}, vol. 26, no. 4, pp. 787--796, 2018.

\bibitem{nachmani2020voice}
Eliya Nachmani, Yossi Adi, and Lior Wolf,
\newblock ``Voice separation with an unknown number of multiple speakers,''
\newblock in {\em ICML}, 2020.

\bibitem{whamr}
Matthew Maciejewski et~al.,
\newblock ``Whamr!: Noisy and reverberant single-channel speech separation,''
\newblock in {\em ICASSP}, 2020.

\bibitem{parallel_wavgan}
Ryuichi Yamamoto et~al.,
\newblock ``Parallel wavegan: A fast waveform generation model based on
  generative adversarial networks with multi-resolution spectrogram,''
\newblock in {\em ICASSP}, 2020.

\bibitem{yamamoto2019probability}
Ryuichi Yamamoto, Eunwoo Song, and Jae-Min Kim,
\newblock ``Probability density distillation with generative adversarial
  networks for high-quality parallel waveform generation,''
\newblock {\em preprint arXiv:1904.04472}, 2019.

\bibitem{WSJ0}
John Garofolo, David Graff, Doug Paul, and David Pallett,
\newblock ``Csr-i (wsj0) complete ldc93s6a,''
\newblock {\em Web Download. Philadelphia: Linguistic Data Consortium}, vol.
  83, 1993.

\bibitem{WHAM}
Gordon Wichern et~al.,
\newblock ``Wham!: Extending speech separation to noisy environments,''
\newblock in {\em Interspeech}, 2019.

\bibitem{rir_generator}
Emanuel~AP Habets,
\newblock ``Room impulse response generator,''
\newblock {\em Technische Universiteit Eindhoven, Tech. Rep}, vol. 2, no. 2.4,
  pp. 1, 2006.

\end{thebibliography}

\end{document}